\documentclass[%
reprint,
amsmath,amssymb,
aps,
longbibliography,
nofootinbib
]{revtex4-1}

\usepackage{hyperref}
\usepackage{graphicx}
\usepackage{color}

\usepackage{bm}
\usepackage{amsmath}
\usepackage{natbib}
\usepackage{textgreek}
\usepackage{mathtools}
\usepackage{filecontents}
\usepackage{footmisc}

\begin{document}
	
	\preprint{APS/123-QED}
	
	\title{Stability of spontaneous, correlated activity in mouse auditory cortex}
	
	\author{Richard F. Betzel$^1$}
	\author{Katherine C. Wood$^2$}
	\author{Christopher Angeloni$^2$}
	\author{Maria Neimark Geffen$^2$}
	\author{Danielle S. Bassett$^{1,3,4,5,6}$}
	\email{dsb @ seas.upenn.edu}
	\affiliation{
		$^1$Department of Bioengineering, School of Engineering and Applied Science, University of Pennsylvania, Philadelphia, PA, 19104 USA}
	\affiliation{
		$^2$Department of Otorhinolaryngology: HNS, University of Pennsylvania, Philadelphia, PA, 19104 USA}
	\affiliation{
		$^3$Department of Electrical and Systems Engineering, School of Engineering and Applied Science, University of Pennsylvania, Philadelphia, PA, 19104 USA}
			\affiliation{
		$^4$Department of Neurology, Perelman School of Medicine, University of Pennsylvania, Philadelphia, PA, 19104 USA}
			\affiliation{
		$^5$Department of Physics \& Astronomy, College of Arts \& Sciences, University of Pennsylvania, Philadelphia, PA, 19104 USA}
		\affiliation{
		$^6$Department of Psychiatry, Perelman School of Medicine, University of Pennsylvania, Philadelphia, PA, 19104 USA}
			\affiliation{
		$^*$To whom correspondence should be addressed: dsb@seas.upenn.edu.}

	\date{\today}
	
	\begin{abstract}
	Neural systems can be modeled as complex networks in which neural elements are represented as nodes linked to one another through structural or functional connections. The resulting network can be analyzed using mathematical tools from network science and graph theory to quantify the system's topological organization and to better understand its function. While the network-based approach has become common in the analysis of large-scale neural systems probed by non-invasive neuroimaging, few studies have used network science to study the organization of biological neuronal networks reconstructed at the cellular level, and thus many very basic and fundamental questions remain unanswered. Here, we used two-photon calcium imaging to record spontaneous activity from the same set of cells in mouse auditory cortex over the course of several weeks. We reconstruct functional networks in which cells are linked to one another by edges weighted according to the maximum lagged correlation of their fluorescence traces. We show that the networks exhibit modular structure across multiple topological scales and that these multi-scale modules unfold as part of a hierarchy. We also show that, on average, network architecture becomes increasingly dissimilar over time, with similarity decaying monotonically with the distance (in time) between sessions. Finally, we show that a small fraction of cells maintain strongly-correlated activity over multiple days, forming a stable temporal core surrounded by a fluctuating and variable periphery. Our work provides a careful methodological blueprint for future studies of spontaneous activity measured by two-photon calcium imaging using cutting-edge computational methods and machine learning algorithms informed by explicit graphical models from network science. The methods are flexible and easily extended to additional datasets, opening the possibility of studying cellular level network organization of neural systems and how that organization is modulated by stimuli or altered in models of disease.
	\end{abstract}

	\maketitle
	
	\section*{Introduction}
	
	Distributed and often redundant coding is a hallmark of neural systems \cite{schneidman2011synergy}, providing robustness to single-neuron variability \cite{montijn2016poplulation} and supporting complexity in the system's potential behavioral repertoire \cite{ganmor2015thesaurus}. A key challenge in understanding this code lies in determining how the nature and strength of correlations between neurons is related to a stimulus \cite{schneidman2003synergy}. Recent evidence suggests that so-called noise correlations have marked and diverse functions \cite{kohn2016correlations}, from impacting information encoding and decoding \cite{averbeck2006effects,bergen2018modeling,ayherabide2013when}, to tuning the amount of information present and thus the nature of ensuing cortical representations \cite{kanitscheider2015origin}. Correlations in spike trains have also been noted to contain important information about excitability, latency, and synchronization \cite{brody1999disambiguating,brody1999correlations,grewe2017synchronous}. Even apart from task-evoked activity, spontaneous activity and correlations of that activity can profoundly impact cortical responses to a sensory input, thereby playing a critical role in information processing \cite{arieli1995coherent, arieli1996dynamics}.
	
	To better understand the nature of coherent multi-unit interactions both during intrinsic processing and during stimulus-induced processing, it is necessary to have a language in which to study inter-unit interaction patterns. In related work in other species and other spatial scales, network science has proven its utility as just such a candidate language \cite{bassett2017network}. The notion of a network in its simplest form is akin to the notion of a graph in the field of mathematics known as graph theory \cite{newman2010networks}. Specifically, an undirected binary graph is composed of nodes, which represent the units of the system, and edges, which link pairs of nodes according to some physical connection, functional relation, or shared feature \cite{butts2009revisiting}. This simplest version of a network can also be expanded to include weights on edges, weights on nodes, dynamics on edges, dynamics on nodes, or multiple types of nodes or edges forming a multilayer or multiplex structure \cite{khambhati2018modeling, vaiana2018multilayer}.  By either the simple or expanded encoding, network models of neural systems seek to distill the most salient organizational features of the system, allowing investigations into how the network topology constrains or supports the system's function \cite{bassett2018nature}. Importantly, the network modeling approach is flexible in the sense that its components can be redefined at different spatial scales, and is thus equally applicable to cellular data at the microscale as it is to regional data at the large scale \cite{betzel2017multi}.
	
	Recent studies have begun to build and characterize network models of cellular activity as measured by calcium imaging \cite{mann2017whole, goltstein2015effects, runfeldt2014acetylcholine, modi2014ca1, warp2012emergence, vanni2017mesoscale, mcvea2016large}, and have demonstrated their biological relevance across a neural system's development. For example, one notable study provided initial evidence that immature cells in the developing brain display spontaneous correlation patterns that are characterized by small-world architecture and that critically regulate neural progenitor proliferation \cite{malmersjo2013small}. In a mature system (ferret visual cortex), recent evidence suggests that local connections in early cortical circuits can generate structured long-range network correlations that guide the formation of visually evoked distributed functional networks that display striking network modularity \cite{smith2018distributed}. The architecture of correlations in spontaneous activity can be regulated by synaptotagmin \cite{chiang2012synaptagmin}, modulated by acetylcholine \cite{runfeldt2014acetylcholine},  blocked by glutamatergic antagonists \cite{mao2001dynamics}, and mediated by a combination of intrinsic and circuit mechanisms \cite{mao2001dynamics}. Yet, little is known about the conservation or variation of network architecture in spontaneous correlations across different regions of the brain. Moreover, while the activity can be temporally quite precise in a given instance \cite{mao2001dynamics}, little is known about how patterns of spontaneous activity change over the course of days and weeks after the critical period of development has passed. Understanding the principles of these dynamics is important for understanding the conserved rules that the architecture must obey, as well as the variability that can be exercised to meet the demands of the ever changing internal or external environments.
	
	Here, we take steps to address some of these gaps in knowledge by measuring correlated spontaneous neuronal activity using two-photon calcium imaging, modeling those correlation patterns as networks, and assessing network architecture and dynamics over the course of several weeks. We focus our measurements specifically on mouse auditory cortex because of its rich organizational characteristics, with distributed representations of tone frequency \cite{funamizu2011distributed}, spatially overlapping locations for the representations of pitch and timbre \cite{allen2017representations}, and the capacity for single neurons within the wider network to encode simultaneous stimuli by switching between activity patterns \cite{groh2018single}. We choose mouse as our species of interest largely to prepare for future efforts using two-photon optogenetics \cite{carrillo2017imaging} to perturb the network architecture, with the goal of probing network response to stimulation and validating recently posited theories of network control \cite{gu2015controllability, yan2017network, kim2018role}. We begin by testing the hypothesis that networks reconstructed from fluorescence correlations exhibit hierarchical modular structure, and that network modules fluctuate over the timescales of days or weeks. We also test the hypothesis that some units participate in these temporal fluctuations more than others such that the system is best characterized by the existence of a stable \emph{temporal core} surrounded by a fluctuating and variable periphery. Each of these hypotheses is motivated by prior observations in non-invasive imaging data acquired from humans \cite{bassett2013task, sporns2016modular, betzel2017positive}, where evidence points to the importance of hierarchical modularity and temporal core-periphery structure for effective cognitive function \cite{sporns2016modular, bassett2017network, bertolero2015modular, gallen2016reconfiguration, arnemann2015functional}. Thus, collectively our hypotheses are predicated on the notion that neural systems are constrained to display some degree of preservation in network architecture across species, from human to mouse \cite{heuvel2016comparative, kim2018role, betzel2018specificity}, as well as scale invariance, from the level of large-scale areas to the level of small-scale units \cite{scholtens2014linking, scholtens2018multimodal}.

	\section*{Results}
	
	We recorded spontaneous activity from four awake, head-fixed mice over the course of 7, 10, 12, and 16 sessions spanning between 2 and 4 weeks. Specifically, we used two-photon microscopy to detect changes in fluorescence of GCaMP6s in transfected neurons caused by fluctuations in calcium activity. We estimated functional connectivity from the fluorescence traces using a cross-correlation of differenced activity for every pair of cells. We modeled the cell-to-cell correlation matrix as a network \cite{bassett2018nature}, and quantitatively characterized the network's architecture using well-developed tools from network science \cite{newman2010networks}. Specifically, we assessed the modularity of the network structure using a commonly applied community detection technique known as modularity maximization \cite{newman2004finding, porter2009communities, fortunato2016community}. Further, we assessed temporal fluctuations in this modular structure using tools for the analysis of dynamic graphs \cite{khambhati2018modeling, sizemore2018dynamic, holme2012temporal}. For further details on our methodological approach, see \textbf{Materials and Methods}.
	
	\subsection*{Networks exhibit multi-scale modular structure}
	
	\begin{figure*}[t]
		\centering
		\includegraphics[width=1\textwidth]{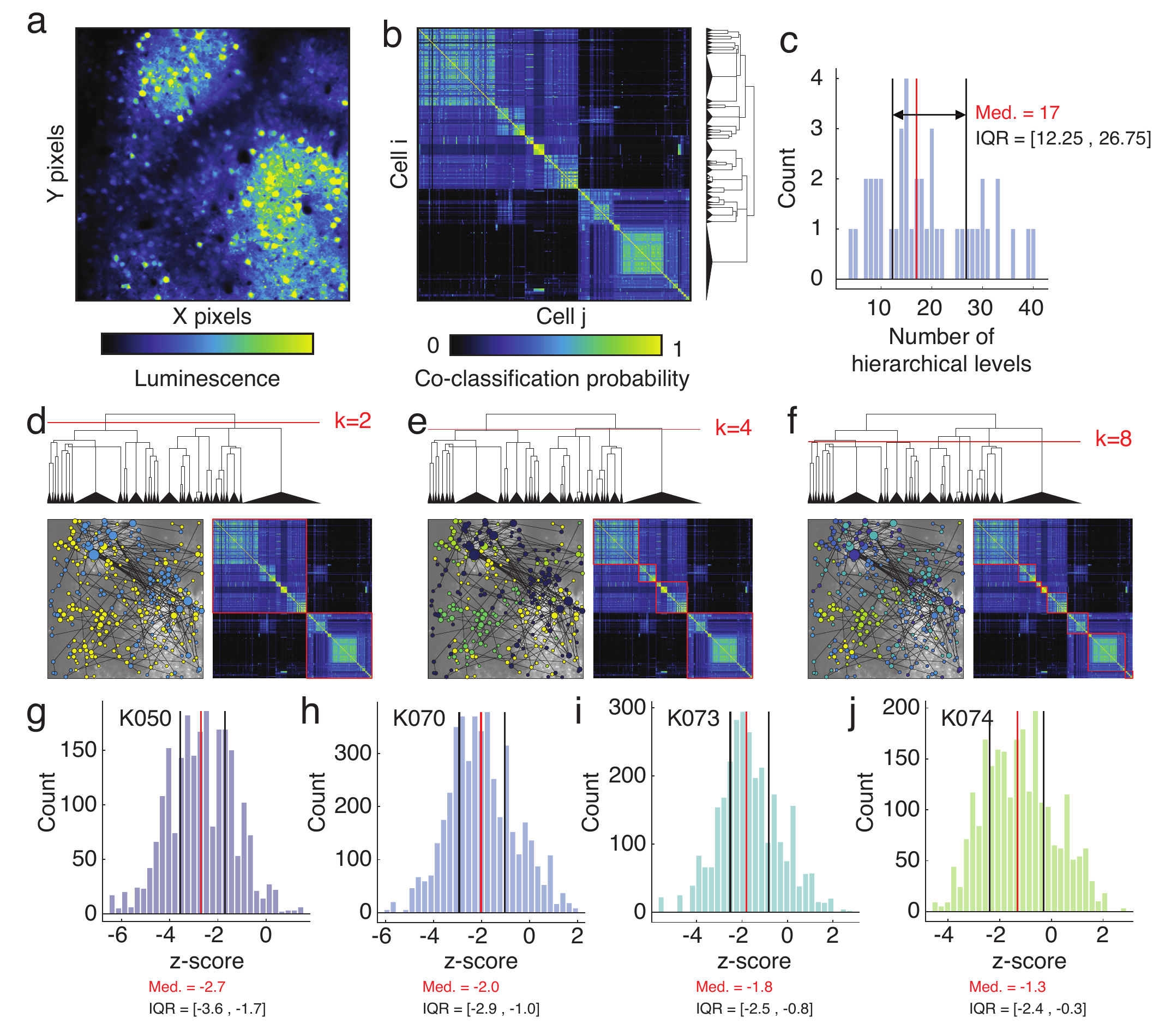}
		\caption{\textbf{Detection of hierarchical modular structure.} (\emph{a}) Mean luminescence of pixels, averaged over the full recording session. (\emph{b}) Co-classification matrix generated using all statistically significant hierarchical levels. The dendrogram to the right depicts module splits. (\emph{c}) The number of hierarchical levels aggregating data from all mice and all recording sessions. Panels (\emph{d}), (\emph{e}), and (\emph{f}) depict module assignments at different levels of the hierarchy. Panels (\emph{g}), (\emph{h}), (\emph{i}), and (\emph{j}) depict distributions of $z$-scored mean intra-module distance for each module and for each mouse. Panels (\emph{b}),(\emph{d}), (\emph{e}), and (\emph{f}) depict representative results from mouse ``K050''.  } \label{hierarchical+modules}
	\end{figure*}
	
	One of the most important organizational principles of biological neural networks is their organization into cohesive modules \cite{sporns2016modular}. These modules are thought to support specialized information processing while conferring robustness to perturbations. Moreover, converging evidence from micro- and macro-scale network analyses suggest that network modules are also organized hierarchically, with larger modules subtending broader brain function and smaller modules playing more specialized roles  \cite{zhou2006hierarchical, meunier2009hierarchical, betzel2017multi}. In this section, we test the hypothesis that networks reconstructed from fluorescence correlations in mouse auditory cortex exhibit hierarchically modular structure.
	
	To address this hypothesis, we leverage recent advances in \emph{community detection methods} \cite{fortunato2010community} -- a collection of algorithms and heuristics that use data-driven approaches to uncover the modular structure of networks. Specifically, we use an extension of the popular \emph{modularity maximization} algorithm \cite{newman2004finding}. The standard version of this algorithm defines a module as a group of network nodes whose internal density of connections is maximally greater than what would be expected under a chance model. The extension of this algorithm samples modules over multiple organization scales, ranging from coarse divisions of the network into a few large modules to finer divisions of the network into many small modules. Importantly, unlike past applications, this extension also includes built-in null statistical testing capable of rejecting modular structure at different levels of the proposed hierarchy if they were consistent with a null model.
	
	Here, we applied this approach to investigate the hierarchically modular structure of networks derived from correlated fluorescence traces (Figure.~\ref{hierarchical+modules}a). The module detection method was applied separately to networks constructed from data in each recording session, which allowed us to take full advantage of all cells recorded on a given day. The algorithm resulted in a hierarchy of communities that survived statistical testing for significance ($p < 0.05$; Figure.~\ref{hierarchical+modules}b--f). In general, we found that the fluorescence networks exhibited hierarchical, multi-scale modular structure. Of the 43 recordings (aggregated across all mice) we observed statistically significant hierarchies in all. Across recording sessions, the average number of scales in a hierarchy was 17 (inter-quartile range of $[12.25,26.75]$; Figure.~\ref{hierarchical+modules}c).
	
	Additionally, we also computed spatial statistics for each module. Past studies have shown that communities tend to be spatially co-localized, so that other cells located near one another are more likely to belong to the same module compared to cells located far from one another \cite{muldoon2013spatially}. To test whether this was also the case in our data, we computed the Euclidean distance from each cell to the nearest cell assigned to the same community. We then averaged this measure over all nodes in the same module. If cells were arranged in spatially dense, compact modules, then this measure would be small. Here, we calculated this measure for each module at every level of the hierarchy and compared these values against a null distribution generated by randomly and uniformly permuting the cell's spatial locations but preserving their module assignments. For each module, we expressed the mean nearest-neighbor distance as a $z$-score with respect to this distribution. We found that the observed modules tended to be more spatially compact than expected by chance. For each mouse, the median $z$-score was less than zero and in all cases the inter-quartile range of $z$-scores excluded a value of zero (Figure.~\ref{hierarchical+modules}g,h,i,j), indicating that the observed modules tended to be more spatially compact than expected by chance.

	\subsection*{Network and module similarity decays over time}

	\begin{figure*}[t]
		\centering
		\includegraphics[width=1\textwidth]{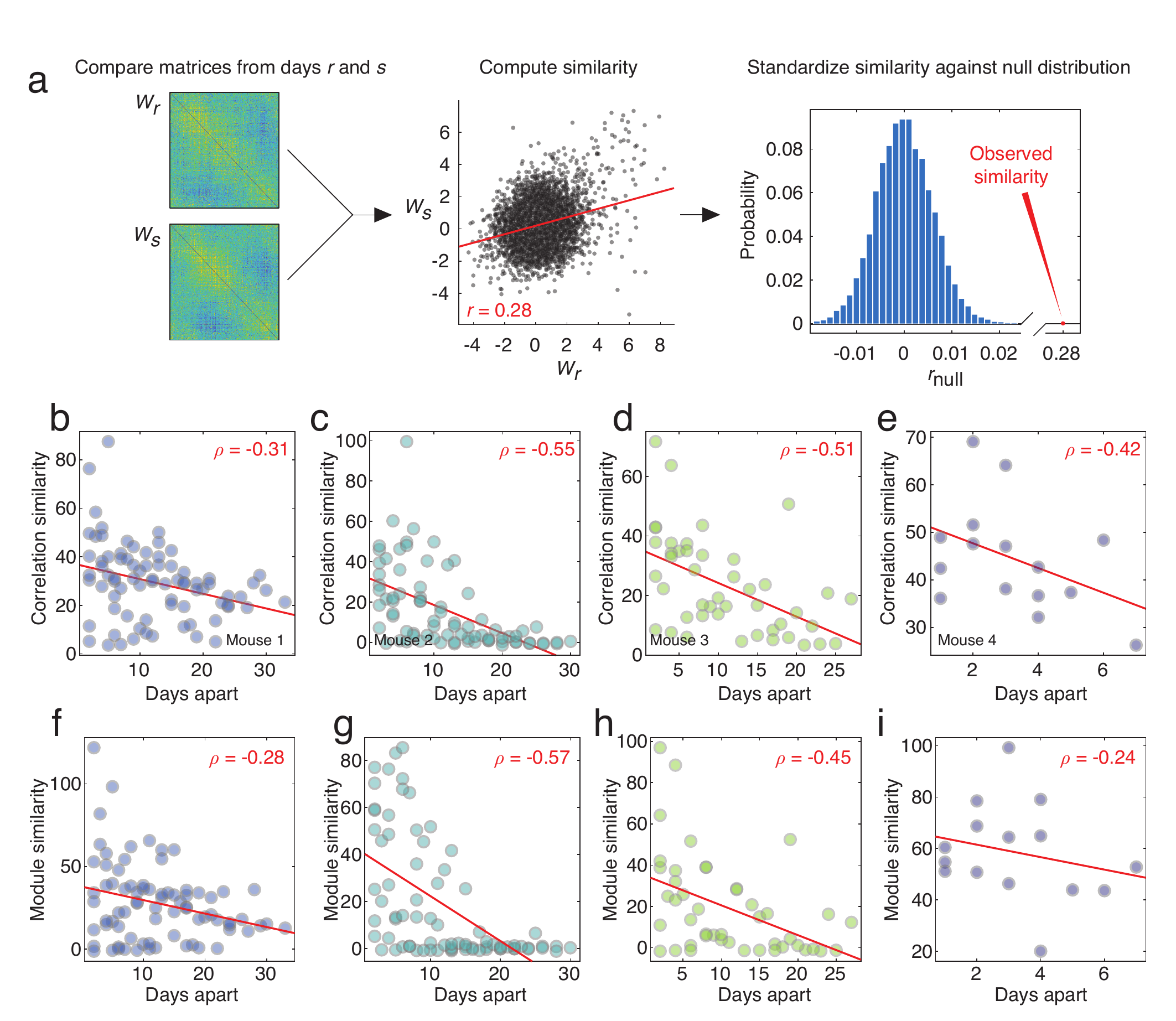}
		\caption{\textbf{Reconfiguration of correlation structure over time.} (\emph{a}) Analysis pipeline for comparing correlation structure. For any two correlation matrices, $W_r$ and $W_s$, whose elements have been $z$-scored following an appropriate jittering procedure (see Methods), we extract the upper triangular elements and compute their similarity using a Pearson correlation coefficient. We compare the observed correlation coefficient against that which we would expect under a null model in which rows and columns of $W_r$ are permuted uniformly at random. In panels (\emph{b}), (\emph{c}), (\emph{d}), and (\emph{e}), we show the scatterplots of standardized similarity scores for pairs of correlation matrices with the number of days separating their respective recording sessions. In panels (\emph{f}), (\emph{g}), (\emph{h}), and (\emph{i}), we show the standardized similarity scores of module co-assignment matrices across pairs of recording sessions.} \label{quotidian+variation}
	\end{figure*}
	
	In the previous section we demonstrated that the correlation pattern of fluorescence traces exhibits modular structure across multiple scales, and that these multi-scale modules unfold as part of a hierarchy. In these analyses, the network's modular structure was derived separately for each recording session. While this approach allowed us to characterize the modular structure on a given day, it tells us little about how those modules fluctuate over the timescales of days or weeks. Here, we address this question directly, taking advantage of the longitudinal tracking of cells across multiple recording sessions to assess the temporal consistency of the network's overall organization, as reflected in the full correlation matrix, and in the network's mesoscale organization, as reflected in its modular structure.
	
	We begin by calculating the similarity between the correlation structure for any two recording sessions, $r$ and $s$ (Figure.~\ref{quotidian+variation}a). We first identified the set of cells from which fluorescence traces were recorded in \emph{both} sessions. We then extracted the correlation structure among those subsets of cells for each of the two recording sessions, resulting in two correlation matrices: $W_r$ and $W_s$. Next, we vectorized the upper triangles of both matrices and computed the correlation of their elements, $\rho_{rs}$. Finally, we expressed this correlation as a $z$-score, $z^W_{rs}$, with respect to a null distribution generated by randomly and uniformly permuting rows and columns of $W_{r}$ and recomputing the correlation of $W_{r}^{\text{perm}}$ with $W_{s}$ (essentially the Mantel test \cite{mantel1967detection}). Accordingly, large positive $z$-scores indicate that the correlation of $\rho_{rs}$ was much greater than expected in the non-parametric permutation-based null model. Aggregating $z$-scores across all pairs of recording sessions resulted in the $z$-scored similarity matrix, $Z^W = [z^W_{rs}]$. 
	
	To assess the degree to which the similarity in correlation structure depended upon the time interval that separated the recordings, we also computed the distance matrix $D = [d_{rs}]$, which measures the distance (in number of days) between recording sessions $r$ and $s$. We then compared the upper triangular elements of $Z^W$ with the corresponding elements of $D$. In general, we observed that $z^W_{rs}$ decayed monotonically as a function of $d_{rs}$. Notably, this observation was consistent across all mice (mean$\pm$standard deviation Spearman rank correlation of $\rho_{z^W_{rs}, d_{rs}} = -0.45 \pm 0.11$) (Figure.~\ref{quotidian+variation}b-e). These findings indicate that the magnitude with which individual cells are correlated with one another over time varies systematically over recording sessions. Specifically, the correlation structures of recording sessions separated by a short period of time tend to be similar to one another, whereas those separated by longer periods of time tend to be dissimilar.
	
	In addition to assessing whether cell-to-cell correlation patterns varied across recording sessions, we also aimed to assess the variability of modular structure. To address this question, we performed an analogous procedure to the one described above where we substitute the module co-assignment matrices $C_r = [C_{ijr}]$ and $C_s = [C_{ijs}]$ for the correlation matrices, $W_r$ and $W_s$. Here, the element $C_{ijr}$ indicates the fraction of all detected community partitions  in which cells $i$ and $j$ were co-assigned to the same module. Otherwise, this procedure for relating the modular structure of networks from different recording sessions proceeded exactly as described above. We denote the $z$-score matrix from the module comparison as $Z^C = [z^C_{ij}]$. As before, we observed that the correlation of similarity in modular structure decays with time (mean$\pm$standard deviation Spearman rank correlation of $\rho_{z^W_{rs}, d_{rs}} = -0.39 \pm 0.15$), indicating the presence of marked quotidian variation (Figure.~\ref{quotidian+variation}f-i).

	\subsection*{Temporal core-periphery structure}
	
	\begin{figure*}[t]
		\centering
		\includegraphics[width=0.8\textwidth]{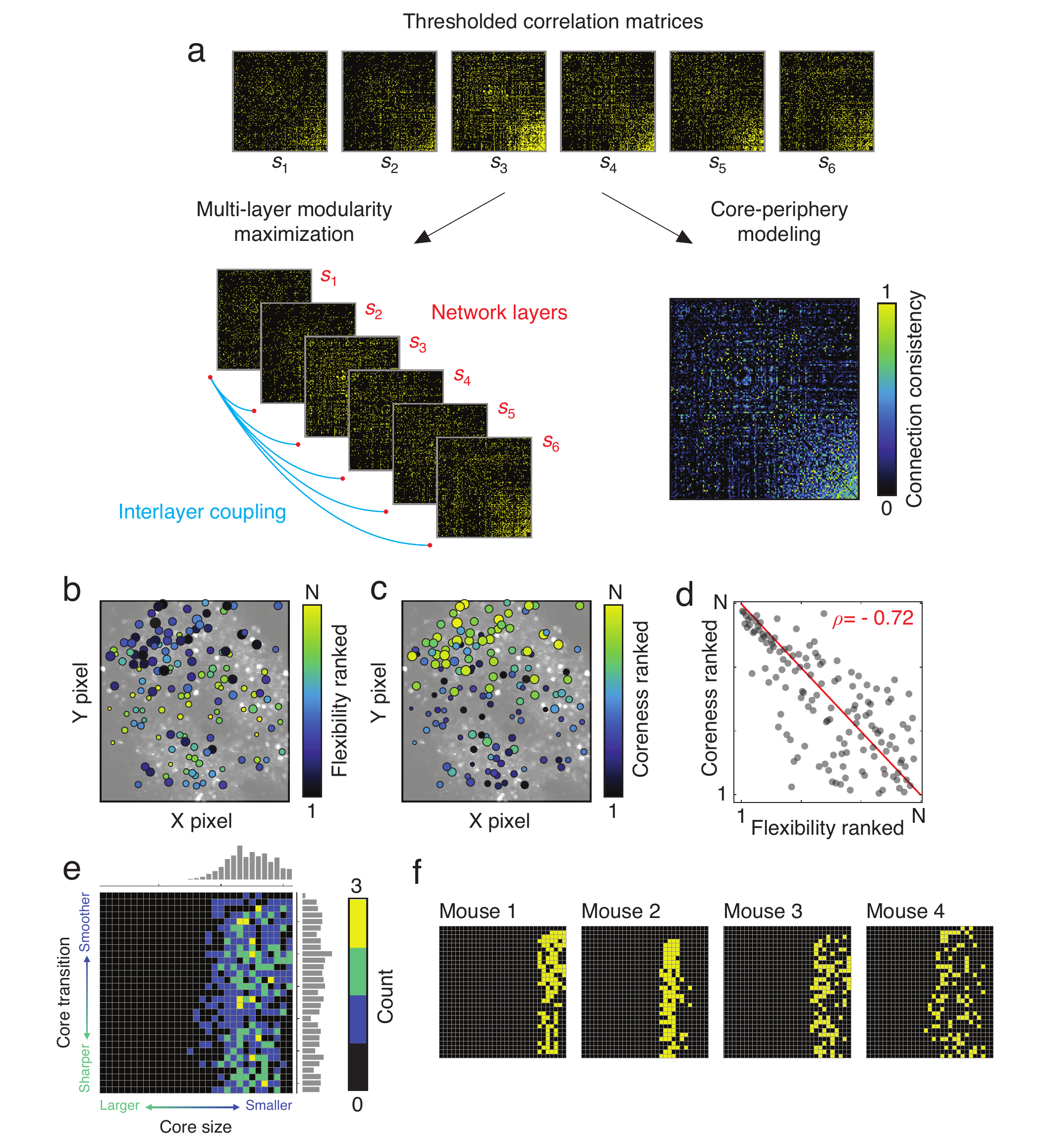}
		\caption{\textbf{Estimation of core-periphery structure and network flexibility.} (\emph{a}) Thresholded correlation matrices are separately treated as: \emph{a}) layers in a multi-layer network, their communities estimated, and network flexibility estimated as the frequency with which a node changes its community assignment across layers; \emph{b}) the consistency matrix is submitted to a core-periphery detection algorithm and each node's ``coreness'' is estimated. (\emph{b}) Node's flexibility scores plotted in anatomical space. (\emph{c}) Nodes' ``coreness'' plotted in anatomical space. The size of nodes in panels \emph{b} and \emph{c} is proportional to their average weight across all six recording sessions. (\emph{d}) Because flexibility is a measure of variability while ``coreness'' is a measure of stability, we find that the two are inversely correlated with one another (red line represents the identity line). (\emph{e}) Cross-subject consistency of optimal parameters for fitting the core-periphery model. For each mouse, we calculated the difference between observed core quality and that of a null model, and we retained the top 10\% of those points. These points are depicted at the level of individual mice in panel \emph{f}. In panel \emph{e}, we aggregate those values across all mice.} \label{core+periphery}
	\end{figure*}

	In the previous section we demonstrated that, on average, hierarchical modular structure becomes increasingly dissimilar over time. However, it may be the case that some sets of brain areas maintain their modular structure despite the passage of time, forming a stable \emph{temporal core} surrounded by a fluctuating and variable periphery \cite{bassett2013task}. To test this hypothesis, we focused on sequences of recording sessions and characterized the stability of modules across those sessions.
	
	Because the number of recording sessions varied from one mouse to another, we focused on sequences of six recording sessions (the greatest number that was available for \emph{all} mice). For each mouse, we modeled the thresholded connectivity data from each of these six recording sessions as the layers in a multi-layer network object \cite{de2017multilayer, vaiana2018multilayer}, and we used multi-layer modularity maximization \cite{mucha2010community} to track the fluctuations in modular structure across those six sessions (Fig.~\ref{core+periphery}a; see \textbf{Materials and Methods} for more details). The multi-layer modularity maximization approach extends the traditional modularity maximization approach \cite{newman2004finding} by detecting modules in all layers (recording sessions) simultaneously. This extension enables us to track the formation and dissolution of modules over time and to seamlessly map modules from one recording session to the next. Given such a mapping, one can then calculate measures like local the ``network flexibility'' \cite{bassett2013robust, bassett2013task}, which indicates how frequently a given node changes its module assignment across layers. Past studies have used this flexibility measure to identify temporally stable cores and variable peripheries (clusters of nodes with low and high flexibility, respectively) \cite{bassett2013task}.
	
	Obtaining estimates of network flexibility requires the detection of communities using multi-layer modularity maximization, which depends upon two parameters, $\gamma$ and $\omega$. These parameters control the resolution (size and number) of modules detected and their stability across layers, respectively. Here, we use a recently developed procedure that allows us to obtain a representative sample from the parameter space defined by these two variables \cite{betzel2018community}. For each such sample, we calculated a local (node-level) measure of flexibility, ranked the flexibility scores of all nodes, and subsequently averaged these ranked flexibility scores across all samples to generate an average flexibility profile for the population of cells.
	
 	In addition to the flexibility approach, we also used a second method to provide converging evidence of temporal core-periphery structure (Fig.~\ref{core+periphery}a) \cite{rombach2017core}. In this procedure, we calculated the session-averaged connectivity matrix (over the six recording sessions), and based on its organization we algorithmically assign cells (nodes) to a continuously defined core and periphery (see \textbf{Materials and Methods} for more details). Intuitively, core nodes are nodes that maintain strong connections to one another and to the periphery across recording sessions, while peripheral nodes are those whose connections are variable (e.g., observed in only a few recording sessions or absent altogether). The size of the core and the smoothness of the transition from core to periphery are controlled by two free parameters, $\alpha$ and $\beta$. We systematically explored this parameter space and at each point, we fit the core-periphery model to the session-averaged network to calculated the core quality \cite{rombach2017core}. We compare the quality of cores fit to the observed session-averaged matrix against the qualities of random matrices generated by a permutation-based null model. This comparison allows us to identify points of interest in the parameter space: points where the observed core was of greater quality than that of the null model.
 	
 	In general, we found evidence that cortical activity in all mice exhibited temporally-stable cores of nodes that maintained community assignments and connectivity over many days. In general, the flexibility measure converged with the coreness measure, implicating roughly the same sets of nodes as temporally stable (i.e., manifesting high coreness, low flexibility) (Fig.~\ref{core+periphery}b-d). Across mice, cores tended to be fairly exclusive (Fig.~\ref{core+periphery}e,f); core quality was maximally greater than the null model at points in parameter space corresponding to a small core. The smoothness of the transition between core and periphery was more variable, suggesting that these networks may exhibit multiple cores with different degrees of smoothness separating the core from the periphery.
	
	\section*{Discussion}
	
	Spontaneous fluctuations in neural activity at the cellular scale can modulate behavioral responses to incoming sensory stimuli \cite{arieli1995coherent,arieli1996dynamics}. Yet the nature of that modulation is not well understood, in part due to the fact that such spontaneous activity does not appear to be random in nature, but instead displays heterogeneous dependencies or correlations among units. Little is known about the rules constraining the architecture of these correlations, or their variability over time. Here we sought to partially address this gap in knowledge by using recently developed techniques in network science to examine the network architecture of correlations in spontaneous activity in mouse auditory cortex as measured by two-photon microscopy and calcium imaging over the course of several weeks. We found that networks exhibited striking modular architecture, with smaller modules being located within larger modules in a multi-scale hierarchy. We also found significant temporal rearrangement of modular architecture, as indicated by the fact that the similarity in modules decreased monotonically as a function of the time interval between recording sessions, even when only considering those units that were present in both sessions. Finally, we found that the broadly observed temporal rearrangement of modules was complemented by the presence of a small number of cells whose modular allegiance remained stable throughout the 2--4 weeks of experimentation. We confirmed with additional testing that the co-existence of stable and unstable units was consistent with a temporal core-periphery model of system dynamics, where a stable core of units is accompanied by a flexibly periphery. Broadly, our study exercises cutting-edge machine learning tools informed by graphical models to study cellular level network organization of neural systems, paving the way for future work examining how that organization might be altered by task demands, developmental stage, or disease burden.

	\subsection*{Multi-scale modular network structure}
	
	Biological systems generally and neural systems specifically, are frequently required to develop, adapt, and evolve in changing environments \cite{kirschner1998evolvability,wagner1996complex}. This pervasive demand for adaptation is thought to be a partial explanation for the striking modular structure observed in biological systems \cite{lipson2002origin,lorenz2011emergence}. Each module is thought to have the capacity to change or adapt without adversely impacting the function of other modules. In neural systems, modules are thought to exist in order to segregate specific cognitive function or computations, allowing enhanced specialization of the organism \cite{sporns2016modular}. Such modular structure has also been observed in spontaneous recordings of intact zebrafish larvae, where topographically compact assemblies of functionally similar neurons reflect the tectal retinotopic map despite being independent of retinal drive \cite{romano2015spontaneous,pietri2017emergence}. These data suggest that spontaneous activity displays modular structure that is a functional adaptation specifically tuned to support the system's behavior. Similar observations have also recently been made in ferret visual cortex, where widespread modular correlation patterns in spontaneous activity accurately predict the local structure of visually evoked orientation columns several millimeters away \cite{smith2018distributed}.
	
	Hierarchical modularity in biological systems is further thought to allow for a decomposability of the system's temporal responses to the environment, with fast processes occurring in small modules at a low level of the hierarchy and slow processes occurring in large modules at a high level of the hierarchy \cite{simon1962architecture}. Prior work at the large-scale has demonstrated the presence of hierarchically modular structure in neural systems specifically, and suggested that large modules support broad cognitive functions while small modules support specialized cognitive functions \cite{zhou2006hierarchical, meunier2009hierarchical, bassett2010efficient, betzel2017multi}. Here we extend these prior observations by showing that over short time periods approximately equal to the duration of a recording session, neurons assemble into cohesive modules of varying size, ranging from large, spatially-distributed clusters of weakly coupled neurons to compact, highly correlated ensembles. In theoretical work, it is interesting to note that hierarchical modularity provides an efficient solution to the problem of evolving adaptable systems while minimizing the cost of connections \cite{mengistu2016evolutionary}. This relation between hierarchical modularity and lost cost yet efficient information processing in neural systems has also been supported by both theoretical work and analysis of neural data in both \emph{C. elegans} and human \cite{bassett2010efficient}. When considering our results in this light, it is useful to note that the spatial compactedness of modules suggests that maintaining long-distance correlated activity may be metabolically costly and therefore uncommon. Overall, these findings are consistent with those observed in other micro- and macro-scale networks and suggest that the organizational principles of modular architecture and spatially-compact, low-cost clusters may be conserved across spatial scales \cite{sporns2016modular,bullmore2012economy}.

	\subsection*{Daily variation in network architecture and module constituency}	
	
	Accompanying the nascent use of tools from network science to understand interaction or connection patterns between neural units, there has been a marked interest in understanding the dynamics of interaction patterns as a function of time, and across a variety of different time scales \cite{kopell2014beyond,calhoun2014chronnectome,khambhati2018modeling,sizemore2018dynamic}. Particularly in the human imaging literature, efforts have begun to understand principles of dynamic network reconfiguration on the time scale of minutes or hours \cite{bassett2011dynamic,xie2018whole}, days \cite{mattar2018network,heitger2012motor}, weeks \cite{bassett2015learning}, months \cite{poldrack2015longterm,betzel2017positive}, and years \cite{martino2014unraveling,zuo2017human}. Here we exercise that interest in the domain of network models of correlation matrices derived from spontaneous activity in mouse auditory cortex over 2 to 4 weeks of experimentation. Our findings suggest that quotidian variation in correlation structure is manifest at multiple scales: (i) at the level of cell-to-cell correlations, but also (ii) at the level of large-scale and module patterns in the network. This latter observation is particularly interesting to consider in light of findings at the macro-scale level of whole-brain networks derived from fMRI data. Specifically, at this large scale, much of the modular organization of spontaneous correlations in the human brain is conserved across the time scales of days and weeks, with notable flexibility largely present at module boundaries. One could speculate that gross temporal stability in macro-scale networks is underpinned by notable micro-scale variability. It would be interesting in future to more directly address the question of the functional role of this micro-scale network reconfiguration, and specifically to test the hypothesis that the correlation structure of fluorescence traces in mouse primary auditory cortex is reorganized over timescales of days to weeks to support cortical functional reorganization. 
	
	\subsection*{A stable network core accompanied by a flexible network periphery}
	
	In other natural dynamical systems, it has been noted that density tends to support temporal stability, while sparsity tends to support temporal instability \cite{huntsman2014density}. In the context of networked systems, the notion can be expanded to describe the phenomenon in which a core of densely interconnected units tends to display weak or slow temporal fluctuations, while a periphery of sparsely interconnected units tends to display strong or fast temporal fluctuations \cite{bassett2013task}. In the context of the human brain, this temporal core-periphery structure has been raised as a model for the balanced constraints of task-general processes, implemented by the temporal core, and task-specific processes, implemented by the temporal periphery \cite{fedorenko2014reworking}. It is interesting to consider whether such a delineation into temporal core and periphery is also characteristic of cellular networks, and whether that separation is functionally meaningful in a similar sense. Our findings suggest that, while calcium fluorescence correlation structure changes markedly over time, there remains a relatively small set of cells whose interactions, both as single connections but also as communities, are spared and preserved. There is some evidence in theoretical studies that such core-like structures emerge early in development, and are strengthened through functional activation \cite{fuchs2009formation}. In analyses of macro-scale networks, core stability and peripheral flexibility have been associated with learning \cite{bassett2013task}, leading us to speculate that the emergence of core-periphery structure in micro-scale networks may serve a similar role in preserving learned (auditory) relationships, while maintaining enough variability to learn and map novel stimuli. Thus, future work could be directed to investigate the functional roles of cores and peripheries during task conditions.

	\subsection*{Methodological Considerations and Limitations}
	
	There are several methodological considerations and limitations that are pertinent to the interpretation and generalizability of our results. First, we note that the experimental methods allow us to sample only a subset of neurons within a specific ``slice'' of the auditory cortex. It is likely that most of the neurons that directly target the neurons that we image are not captured by the analysis. Therefore, the estimates for the network connectivity should not be taken as an approximation for the actual physical connectivity in the cortical circuit. Another important aspect of data collection is that we focus on a specific cortical layer: layer 2/3. Neurons in the cortex differ tremendously in their connectivity patterns across different layers \cite{llano2009differences,theyel2010specific}. It would be important in future studies to sample the activity across cortical depth to better understand integration of information across cortex. 
	
	Second, we note that we have examined correlations in spontaneous activity fluorescence traces, and this approach has the strengths of computational simplicity and ease of interpretation \cite{zalesky2012use}. However, we acknowledge that correlation-based approaches focus on pairwise functional interactions, and remaining agnostic to underlying structural connectivity as well as to higher-order (non-pairwise) relations between units. It would be interesting in future to consider maximum entropy models as an alternative method to estimate connections between units \cite{onken2012maximum}, both for its sensitivity to underlying structure \cite{watanabe2013pairwise}, and for its ability to assess higher-order interactions \cite{ganmor2011sparse}. Approaches that could then take advantage of the richer assessment of higher order interactions in these data include emerging tools from algebraic topology \cite{giusti2016twos,sizemore2018importance}, which have already proven relevant for understanding structure-function relationships at both large and small scales in neural systems \cite{sizemore2018cliques,reimann2017cliques}. 
	
	Finally, an additional limitation concerns the measures used to establish the presence or absence of connections between cells. Specifically, we constructed networks where nodes represented individual cells and where edges represented the correlation magnitude of fluorescence traces. Importantly, correlated activity is not a direct proxy for underlying structural connectivity \cite{goni2014resting}, and thus a pair of neurons that may not be directly synaptically connected can exhibit correlated activity; rather than reflecting structural connections, functional connectivity provides information about the interactions between neurons due to their function \cite{bassett2017network}. Moreover, correlated activity also does not represent the coupling matrix that prescribes the temporal evolution of brain activity \cite{cole2016activity}. Rather, the correlation structure of neural activity represents the product of a dynamical system whose evolution is constrained by structural connections. Though correlated activity at the large-scale has proven useful for investigating the functional organization of brain networks  \cite{vanni2017mesoscale, craddock2013imaging, horwitz2003elusive}, its utility for understanding and characterizing the structure and function of micro-scale networks remains unclear and largely untested \cite{schroter2017micro}. Future studies should both investigate in greater detail the relative advantages of alternative, domain-specific measures of functional connectivity \cite{stetter2012model, feldt2011dissecting} and the relationship of these measures to other connection modalities \cite{scholtens2018cross}.
	
	\subsection*{Conclusion}
	
	Across many scientific disciplines from plant biology \cite{duran2017bridging} to biogeodynamics \cite{zerkle2018biogeodynamics} and the study of biodiversity \cite{hirt2018bridging}, scientists are faced with the challenge of bridging two or more scales of investigation into the function of complex systems. For example, in evolutionary biology, a key challenge is to bridge physical scales from protein sequences to fitness of organisms and populations \cite{bershtein2017bridging}, while in the study of cancer progression a key challenge is to map genotype to phenotype \cite{gerlee2015bridging}. Neuroscience is no exception. Ongoing efforts seek to bridge the gap between the connectome and the transcriptome \cite{fornito2018bridging}, between brains and social groups \cite{falk2017brain}, or between large-scale brain regions and small-scale cellular circuitry \cite{towlson2018bridging}. In each case, the development of a formal understanding will depend upon the capacity to build mathematical descriptions and theories across scales. One natural approach to this challenge is to use a formalism that is scale invariant, a characteristic that makes network science particularly appealing. Our work in this study is an example of considering tools and conceptual paradigms previously exercised at the large-scale of brain regions, and exercising them at the level of cellular circuitry. We look forward to future efforts explicitly measuring and examining the network architecture of neural systems across both of these scales simultaneously in the same animal, with the goal of better understanding and predicting behavior.

	\section*{Methods}
	
	\subsection*{Animals}
	
All experiments were performed with equal numbers of adult male and female mice (supplier -- Jackson Laboratories; age, 12-22 weeks; weight, 20-36 g; PV-Cre mice, strain: B6;129P2-Pvalbtm1(cre)Arbr/J). All experimental procedures were performed in accordance with NIH guidelines and approved by the IACUC at the University of Pennsylvania. 

\subsection*{Two-photon microscopy and calcium imaging}

Four mice were implanted with cranial windows over auditory cortex. Briefly, the mice were anaesthetized with 1.5-3\% isoflurane and a 3mm circular craniotomy was performed over auditory cortex (stereotaxic coordinates) using a 3mm biopsy punch. An adeno-associated virus (AAV) vector encoding the calcium indicator GCaMP6s (AAV1-SYN-GCAMP6s, UPENN vector core) was injected for expression in layer 2/3 neurons in left A1 within the window (750 nl, 1.89 $\times$ 10-12 genome copies per ml) \cite{chen2013ultrasensitive}. After injection a glass circular 3mm coverslip (size 0, Warner Instruments) was placed in the craniotomy and fixed in place using a mix of Krazy glue and dental cement. A custom-made stainless steel head-plate (eMachine Shop) was fixed to the skull using C\&B Metabond dental cement (Parkell). All imaging sessions were carried out inside a single-walled acoustic isolation booth (Industrial Acoustics) as previously described. Mice were placed in the imaging setup, and the headpost was secured to a custom base (eMachine Shop) serving to immobilize the head. Mice were gradually habituated to the apparatus over 3 days, 3-4 weeks after surgery. 

Using two-photon microscopy (Ultima in vivo multiphoton microscope, Bruker) changes in fluorescence of GCaMP6s in transfected neurons caused by fluctuations in calcium activity were recorded in awake, head-fixed mice. We recorded from the same cells over many days in layer 2/3 of auditory cortex, using blood vessel architecture, depth from the surface, and the shape of cells to return to the same imaging site. Laser power at the brain surface was kept below 30 mW. Chronic imaging of the same field of view across days was carried out for the duration of the experiment. 

Recordings were made at 512$\times$512 pixels and 13-bit resolution at approximately 30 frames per second. Spontaneous activity was recorded for 10 minutes in each session. Publicly available toolboxes \cite{pachitariu2017suite} were used to register the resulting images, select regions of interest, estimate neuropil contamination, and extract the changes in fluorescence from each cell. Upon conclusion of the imaging sessions, brains were extracted following perfusion in 0.01M phosphate buffer pH 7.4 (PBS) and 4\% paraformaldehyde (PFA), post-fixed in PFA overnight and cryopreserved in 30\% sucrose solution for 2 days prior to slicing. The location and spread of GCaMP6s was confirmed through fluorescent imaging. These methods are consistent with the recommendations of the American Veterinary Medical Association (AVMA) Guidelines on Euthanasia.
	
	\subsection*{Network reconstruction}
	
	We estimated functional connectivity from fluorescence traces. Let $x_i(t)$ indicate the intensity of fluorescence in cell $i$ at time $t$. Next, we computed the cross-correlation of fluorescence traces for every pair of cells:
	
	\begin{equation}
	W_{ij} = \frac{\sum_{t} (x_i(t) - \mu_i) (x_j(t) - \mu_j) }{\sigma_i\sigma_j}
	\end{equation}
	
	\noindent where $\mu_i$ and $\sigma_i$ are the mean and standard deviation of the differenced time series. To reduce the likelihood that the observed correlations were driven by chance fluctuations, we ``jittered'' cells' time series (by adding or subtracting $<1$ second offsets), and computed jittered cross correlations, $W_{ij}^{jitter}$. We repeated this procedure 1000 times. We estimated for every pair of cells the probability that the jittering procedure would generate a correlation as strong as that which was observed empirically, and we made binary connections between those cells with $p < 0.05$. This procedure resulted in a sparse matrix, $A \in \mathbb{R}^{N \times N}$ with elements $A_{ij} \in [0,1]$.
	
	\subsection*{Module detection}
	We used modularity maximization to detect network modules based on connectivity data \cite{newman2004finding}. This method aims to divide network nodes (cells) into modules whose internal density of connections is maximally greater than what would be expected under a null model. This intuition is formalized by the modularity quality function \cite{reichardt2006statistical}:
	
	\begin{equation}
	Q(\gamma) = \sum_{ij} \bigg[ A_{ij} - \gamma \frac{k_ik_j}{2m} \bigg] \delta(g_i,g_j).
	\end{equation}
	
	In this equation, $k_i = \sum_j A_{ij}$ is the degree of node $i$. The term $\frac{k_ik_j}{2m}$ gives the expected number of connections between node $i$ and node $j$ given the null model in which each node's degree is preserved but connections are formed at random. The resolution parameter, $\gamma$, scales the relative contribution of the null model. The module assignment of node $i$ is encoded as $g_i$ and $\delta(g_i,g_j)$ is the Kronecker delta, whose value is equal to unity when $g_i = g_j$ and is zero otherwise.
	
	In this manuscript, we used two variants of modularity maximization. First, we studied the network community structure for each recording session independently. For this analysis, we combined modularity maximization with a newly-developed multi-resolution technique that divides the network into communities of different sizes (scales) that are related to one another hierarchically \cite{jeub2018multiresolution}. This procedure allows us to examine community structure across a range of scales, from large communities to smaller communities that might support more specialized information processing.
	
	Additionally, we used a multi-layer variant of modularity maximization that makes it possible to track the evolution, formation, and dissolution of communities across recording sessions \cite{mucha2010community}. In this procedure, the standard modularity maximization equation is modified to read:
	
	\begin{equation}
	Q(\gamma, \omega) = \sum_{ijsr} [(A_{ijs} - \gamma k_{is} k_{js})]\delta (g_{is},g_{js}) + \delta (i,j) \cdot \omega ]\delta (g_{is}, g_{jr}).
	\end{equation}
	
	\noindent Here, the subscript $s$ denotes network layers, $s \in \{1 , \ldots , T \}$. So $A_{ijs}$ represents the presence or absence and weight of the connection between node $i$ and node $j$ in layer $s$. Similarly, $k_{is} = \sum_j A_{ijs}$ is the degree of node $i$ in layer $s$ and $g_{is}$ is the community to which node $i$ is assigned in layer $s$. Unique to the multi-layer variant of modularity maximization is the \emph{inter-layer coupling parameter} $\omega$, which links node $i$ to itself across layers. From the perspective of maximizing $Q$, non-zero values of $\omega$ make it advantageous to group node $i$ into the same community across layers. When $\omega$ is small, the advantage is correspondingly small, and the detected communities emphasize the unique community structure of layers. On the other hand, when $\omega$ is large, the detected communities are consistent across layers and emphasize shared features of community structure.
	
	Here, we used a recently-developed procedure to obtain estimates of community structure with the values of $\{ \gamma, \omega \}$ sampled from a restricted parameter space \cite{betzel2018community}. This procedure involved first estimating the boundaries of a restricted parameter space wherein any $\{  \gamma , \omega \}$ pair would result in community structure where the number of communities is $>1$ and $<N \times T$ (where $T$ is the total number of layers; $T = 6$, in this case), and where community structure is neither uniform across layers (flexibility of exactly 0) nor is it maximally dissimilar (flexibility of exactly 1). See \cite{betzel2017multi} for more details on how these boundaries were estimated. We then sampled 10000 $\{  \gamma , \omega \}$ pairs from within this parameter space and for each sample we maximized the corresponding $Q(\gamma,\omega$. All subsequent analyses were carried out on these detected communities.
	
	The principle advantage of the multi-layer formulation is that it estimates communities for all layers simultaneously and preserves nodes' community labels across layers. This advantage makes it possible to directly compare the community assignment of a given node in layer $s$ and in layer $t \ne s$, and to identify nodes whose community assignments are flexible (varying from one layer to another) or inflexible (remaining in the same community across layers). We can quantify this intuition using the network measure \emph{flexibility} \cite{bassett2011dynamic,bassett2013robust}:
	
	\begin{equation}
	f_i = 1 - \frac{1}{T - 1}\sum_{s = 1}^{T - 1} \delta (g_{i,s} , g_{i,s + 1}).
	\end{equation}
	
	\noindent Intuitively, flexibility counts the fraction of times that nodes' community assignments in layers $s$ and $s + 1$ differ. Nodes that differ more frequently have flexibility values closer to 1, while nodes that differ less frequently have flexibility values closer to 0. Here, we used the flexibility measure as an index of change in network community structure across recording sessions.
	
	\subsection*{Core-periphery detection}
	
	Separately, we also characterized the stability of network organization across recording sessions by computing a temporal core and periphery. In this context, a \emph{core} refers to a group of nodes that are densely internally connected and to the \emph{periphery}, which is weakly internally connected \cite{borgatti2000models}. To identify temporal core-periphery structure, we first generated a connection consistency matrix, whose element $G_{ij} = \sum_s A_{ijs}$ represented the fraction of layers (recording sessions) in which a network connection was present. In this matrix, a core refers to a group of nodes whose connections are maintained across time, while the periphery is a set of nodes whose connections are more variable.
	
	We used a variant of a common core-periphery definition in which the transition from core to periphery varies smoothly (non-binary). We begin by defining the $N \times 1$ vector $C_i$ of non-negative elements \cite{rombach2017core}. Given this vector, we then defined the matrix $C_{ij} = C_i C_j$ subject to the constraint that $\sum_{ij} C_{ij} = 1$. The values in the vector $C$ are permutations of the vector:
	
	\begin{equation}
	C_m^* = \frac{1}{1 + exp(-(m - \beta N) \times tan ( \pi \alpha /2 ) )}.
	\end{equation}
	
	\noindent The coreness of each node is the permutation of $C_m^*$ that maximizes the core quality function:
	
	\begin{equation}
	R = \sum_{ij} G_{ij} C_i C_j.
 	\end{equation}
 	
 	\noindent This method introduces two free parameters, $\alpha \in [0,1]$ and $\beta \in [0,1]$. The value of $\alpha$ determines the sharpness of the core-periphery boundary. With $\alpha = 1$, the transition is binary while the transition with $\alpha = 0$ is maximally fuzzy. Similarly, the value of $\beta$ determines the size of the core; as $\beta$ ranges from 0 to 1, the size of the core varies from $N$ to 0. In our application, we performed a grid search of 31 linearly-spaced values of $\alpha$ and $\beta$, using a simulated annealing algorithm to maximize $R$ (with 10 restarts).
	
	\section*{Author Contributions}
	
	MNG raised financial support for the experiments and directs the laboratory in which the experiments were performed. MNG and KCW designed the experiments and fluorescent signal analysis. KCW performed the experiments and provided expertise on data preprocessing and cleaning. CA assisted with experimental data collection. RFB performed the numerical experiments, mathematical modeling, and computational analysis, constructed the figures, and drafted the paper. DSB obtained financial support for the computational and theoretical work, and directs the laboratory in which the computation and theory were performed. DSB also drafted the paper. All authors revised the paper and approved its final contents.

	\section*{Acknowledgements}
	
	RFB and DSB would like to acknowledge support from the John D. and Catherine T. MacArthur Foundation, the Alfred P. Sloan Foundation, the Army Research Laboratory and the Army Research Office through contract numbers W911NF-10-2-0022 and W911NF-14-1-0679, the National Institute of Health (2-R01-DC-009209-11, 1R01HD086888-01, R01-MH107235, R01-MH107703, R01MH109520, 1R01NS099348 and R21-M MH-106799), the Office of Naval Research, and the National Science Foundation (BCS-1441502, CAREER PHY-1554488, BCS-1631550, and CNS-1626008). MNG and KW acknowledge the support of Human Frontier in Science Foundation Young Investigator Award; National Institutes of Health (Grant numbers NIH R01DC014700, NIH R01DC015527), and the Pennsylvania Lions Club Hearing Research Fellowship to MGN. The content is solely the responsibility of the authors and does not necessarily represent the official views of any of the funding agencies.

%	\section*{Code}
	
%	\clearpage
	\newpage
	\bibliographystyle{naturemag}
	\bibliography{bibfile,bibfile_dsb}
	
\end{document}